\documentclass[aps,prl,twocolumn,groupedaddress]{revtex4-1}
\usepackage{graphics}
\usepackage{amsmath}
\usepackage{graphicx}
\usepackage{epstopdf}
\graphicspath{./fig/}

\begin{document}

\title{Emergent low-energy bound states in the two-orbital Hubbard model}

\author{Y. N\'u\~nez-Fern\'andez$^1$}
\email[]{yurielnf@gmail.com}
\author{G. Kotliar$^2$}
\author{K. Hallberg$^1$}
\affiliation{$^1$ Centro At{\'o}mico Bariloche and Instituto Balseiro, CNEA and CONICET, 8400 Bariloche, Argentina}
\affiliation{$^2$ Department of Physics and Astronomy, Rutgers University, Piscataway, New Jersey, 08854, USA}

\date{\today}

\begin{abstract}

A repulsive Coulomb interaction between electrons in different orbitals in correlated materials can give rise to bound quasiparticle states. 
We study the non-hybridized two-orbital Hubbard model with intra (inter)-orbital interaction $U$ ($U_{12}$) and different band widths using an improved dynamical mean field theory numerical technique which leads to reliable spectra on the real energy axis directly at zero temperature. 
We find that a finite density of states at the Fermi energy in one band is correlated with the emergence of well defined quasiparticle states at excited energies $\Delta=U-U_{12}$ in the other band. These excitations are inter-band holon-doublon bound states. At the symmetric point $U=U_{12}$, the quasiparticle peaks are located at the Fermi energy, leading to a simultaneous and continuous Mott transition settling a long-standing controversy.

\end{abstract}


\maketitle

{\it Introduction}: 
The correlation-driven (Mott) metal-insulator transition in interacting materials is one of the central problems in condensed matter theory.  Significant advances in our of  understanding of this phenomena, have taken place over the last decade  for one-band models, and  complete  solutions of this model are now available in the limits of one and infinite dimensions, and in many regimes of the two dimensional case.    The  multi-orbital case, which is more relevant to most materials but at the same time  much more complex,   is still  a very open problem, and even qualitative understanding of many issues  lacking.   One case in point is the study of multi-orbital models with different bandwidths,  and in particular, to study the orbital selective Mott transition (OSMT), in which one band locks in an insulating state, while the other (with a larger bandwidth) remains metallic.
 
Recent theoretical studies of  the simplest model Hamiltonians relevant to  the OSMT phenomena, for example, the two-orbital Kanamori-Hubbard (KH) model which includes a ferromagnetic Hund coupling $J$ between the orbitals, indeed find an OSMT even in the presence of interband hybridization and crystal field splitting.\cite{demedici,demedici1,demediciHund,sigrist,Ferrero2005} However, contradictory results remain for the symmetric case when $J=0$: While some results showed that for this case there is no OSMT for any bandwidth ratio \cite{Liebsch,Koga,Cornaglia}, others claimed that such a transition would take place for a bandwidth ratio of 1/5. \cite{Ferrero2005,demedicigeorgesbiermann,demedici}

Signatures of orbital-selective Mott physics  are ubiquitous in correlated materials. They can be  found in 4$f$ and 5$f$ rare earth heavy fermion materials such as CeRhIn5,  CeCu$_{6x}$Au$_x$, YbRh$_2$Si$_2$, uranium compounds\cite{Vojta}, uranium oxides\cite{huang,lanata}, 
iron pnictide superconductors\cite{Hosono,demedicigiovanetti,yi,miao}, transition metal oxides such as manganites La$_{1-x}$Sr$_x$MnO$_3$ \cite{jaime}, VO$_2$\cite{VO2}, Ca-doped Sr$_2$RuO$_4$ \cite{Anisimov,Ca} and, interestingly, $^3$He bilayers.\cite{He}

In this paper we use a state-of-the-art numerical method based  DMFT using o DMRG as the impurity solver  to study the simplest version of the two-orbital KH model with different bandwidths, considering a finite inter-orbital repulsive Colomb interaction $U_{12}\le U$ even if $J=0$ . 

By calculating the paramagnetic density of states (DOS) for the half-filled case at zero temperature, we find conspicuous quasiparticle (QP) peaks at energies $U-U_{12}$ for the metallic phase. When the OSMT takes place the quasiparticle peaks remain as in-gap states within the Mott gap in the narrower insulating band, while vanishing from the metallic band. For larger interactions, in the insulating phase,  all in-gap states disappear. We conclude that the QP bound states exist in one band only in the presence of a coherent metallic resonance in the other band.
 These quasiparticles are mainly formed by inter-orbital holon-doublon pairs, carrying an energy  $U-U_{12}$. 

An important consequence is that the presence of the QPs implies that when $U=U_{12}$ there is no OSM transition for any non-zero band width in both bands, and the narrow band remains metallic even for an infinitesimal bandwidth, as long as the wide band is metallic. With this result we settle the controversy mentioned above.\\ \\

{\it Model and Method}: We consider the following model Hamiltonian for two interacting orbitals:
\begin{eqnarray}
\label{eq:Hred}
H&=&\sum_{\langle ij\rangle \alpha \sigma}t_{\alpha}c_{i\alpha\sigma}^{\dagger}c_{j\alpha\sigma}+\\ \nonumber
&+&U\sum_{i\alpha}n_{i\alpha\uparrow}n_{i\alpha\downarrow}+\sum_{i\sigma\sigma'}U_{12}n_{i1\sigma}n_{i2\sigma'}
\end{eqnarray}
where $\langle ij\rangle $ are nearest-neighbor sites on a Bethe lattice, $c_{i\alpha \sigma}^{\dagger}$ creates an electron at site $i$, orbital  $\alpha=1,2$ with spin $\sigma$ and $n_{i\alpha }=n_{i\alpha \uparrow} + n_{i\alpha \downarrow}$. $U$ ($U_{12}$) is the intra (inter)-orbital Coulomb repulsion between electrons. The nearest neighbor hoppings are $t_{1} \ge t_{2}$, for the wide (WB) and narrow (NB) bands respectively. We set $t_{1}=0.5$ as the unit of energy and we define $\Delta=U-U_{12}$.

We solve this Hamiltonian at half filling (electron-hole symmetric) using the DMFT\cite{review} and using the DMRG\cite{white1, karen1} to solve the impurity's Green's functions. \cite{garcia,EPLreview}. This allows us to obtain the DOS  directly on the real axis (or with a very small imaginary offset $0.01< \eta <0.2$ \cite{pinning}), zero temperature and system sizes of $L=40-60$ sites.

\begin{figure}
 \includegraphics[scale=0.5]{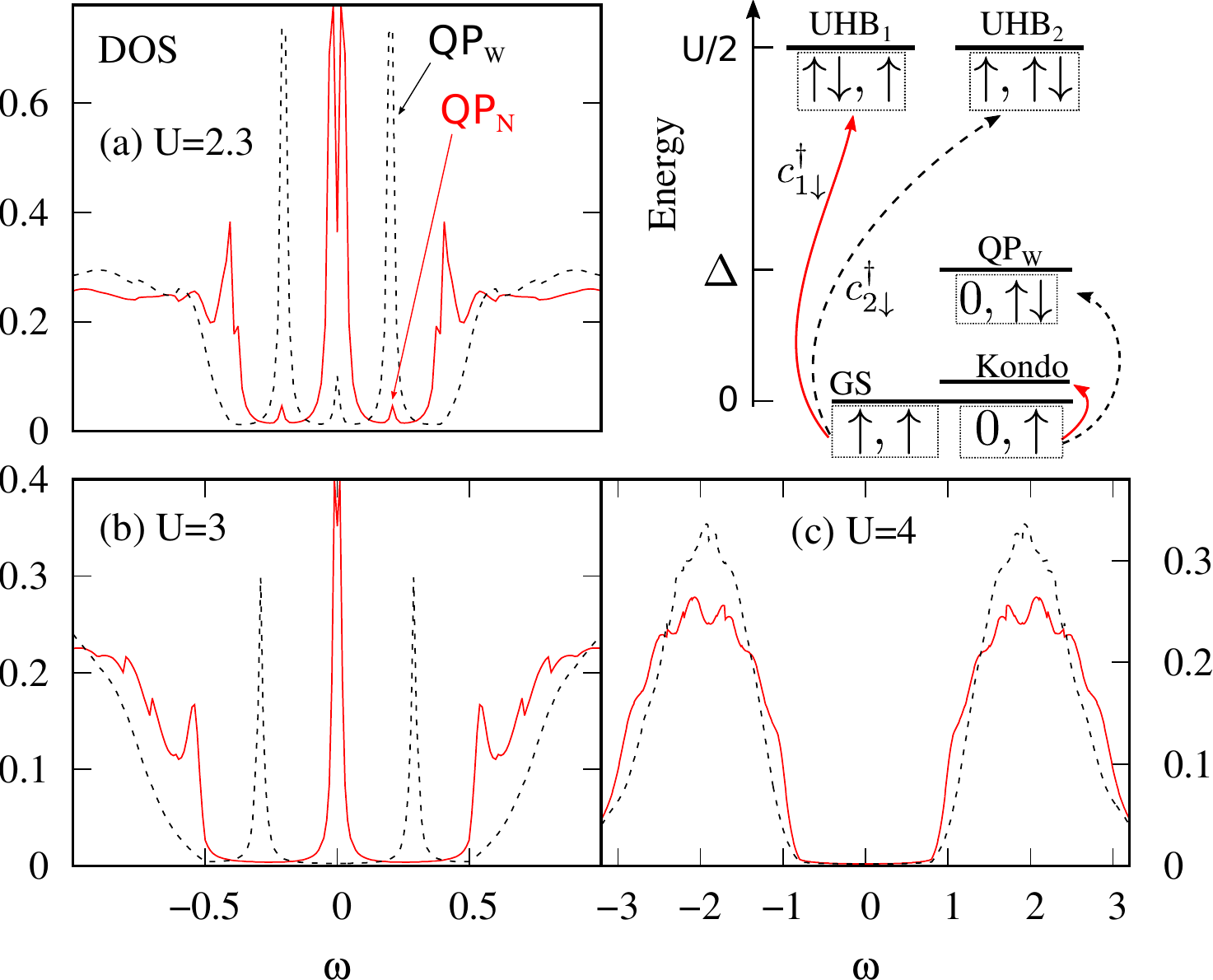}
 \caption{(color on line) DOS showing the OSMT as a function of $U$ for $\Delta=U-U_{12}=0.3$, $t_2=t_1/2$.(a) fully metallic  (here $\Delta=0.2$ for visual purposes) where both quasiparticles QP$_W$ and  QP$_N$ are shown, (b) OSM phase retaining only QP$_W$, and (c) fully insulating states (note the energy axis difference). Red (continuous line): WB; black (broken line): NB.
The sketch shows the relevant states of the effective impurity $|$WB,NB$\rangle $ (omitting the bath sites, for clarity) participating in each main feature of the DOS of the OSM phase (b): the central Kondo peak, the in-gap QP peaks and the Hubbard Bands (only positive energies are shown). Single up spins can be also down spins (see Supp. Mat.).}
\label{figure1}
 \end{figure}

{\it Results}: 
In Fig. \ref{figure1} we present the results for the DOS close to the Fermi energy showing the OSMT in the narrow band as an intermediate phase between the fully metallic and the fully insulating phases with increasing $U$.

A new striking feature is the structure in the form of well defined peaks at energy $\Delta$ from the central coherent resonance present in both bands in the metallic regime (see also Fig. \ref{figure3}).
As described below, we identify these peaks as two new quasiparticles (QP) present in the system: Holes and doublons in the WB (NB) generate wide (narrow) QP states QP$_W$ (QP$_N$) seen as peaks in the narrow (wide) band.
When the interaction is increased and the system goes through an OSMT, the insulating band retains the QP peaks while, obviously, losing the central peak. As a consequence, the QP$_N$ excitations vanish as seen by the disappearence of the QP peaks in the metallic band. For large interactions, when the whole system turns insulating, all in-gap structure disappears, including the QP peaks. The low-energy QPs in one band have the same origin as the central peak in the other band. We also find a suggestive correlation between these peaks: The metallic peak weight is approximately twice the weight of each corresponding QP peak in the other band (see Fig.\ref{figure1}(a); the splitting of the central peak is a finite size effect). This is another indication of their correspondence. When one band is close to the Mott transition its central quasiparticle weight decreases and its corresponding QPs in the other band get heavier, until they both vanish at the transition. These QP exist also for equal bandwidths and they vanish simultaneously at the Mott transition.

\begin{figure}
 \includegraphics[scale=0.5]{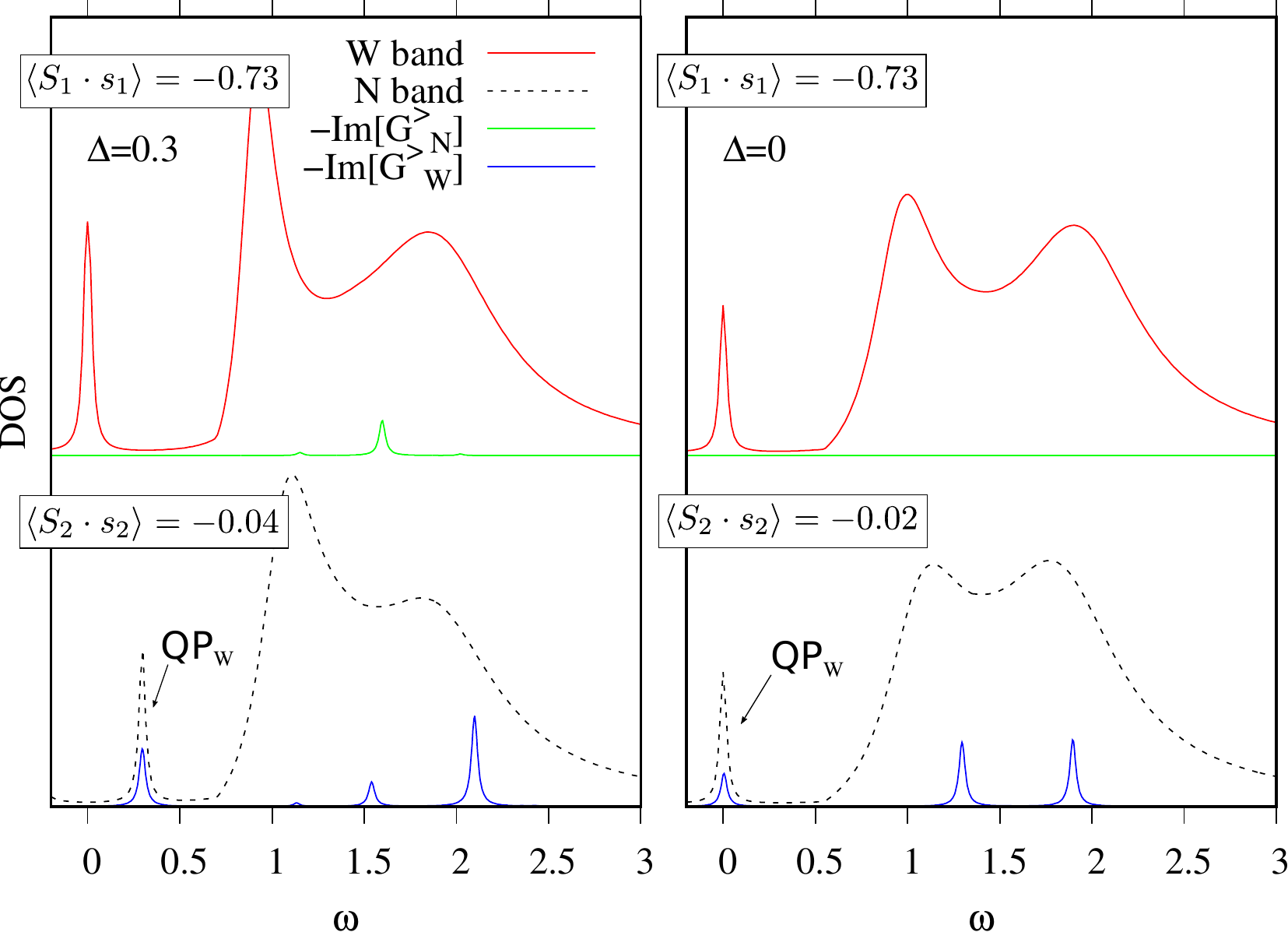}
 \caption{\label{figure2} Density of states of a system with $L=8$ sites  (3 bath sites per impurity orbital) calculated using exact diagonalization. The blue (green) curves correspond to the excitation spectra of $G_{W}^>$ ($G_{N}^>$, see text) representing the holon-doublon pair between bands. The QP peaks correspond to these excitations, which exist also within the Hubbard bands.  Left: $U=3$, $\Delta=0.3$ and $t_2=0.5t_1$. Right:  $U=3.1$, $\Delta=0$ and $t_2=0.02t_1$. We also show the spin-spin correlation in the ground state $\langle {\bf S}_{\alpha}\cdot {\bf s}_{bath\alpha }\rangle$ showing that the central peak in the wide band is Kondo-like, while the one in the narrow band is of the holon-doublon kind. } 
 \end{figure}

\begin{figure}
 \includegraphics[scale=0.45]{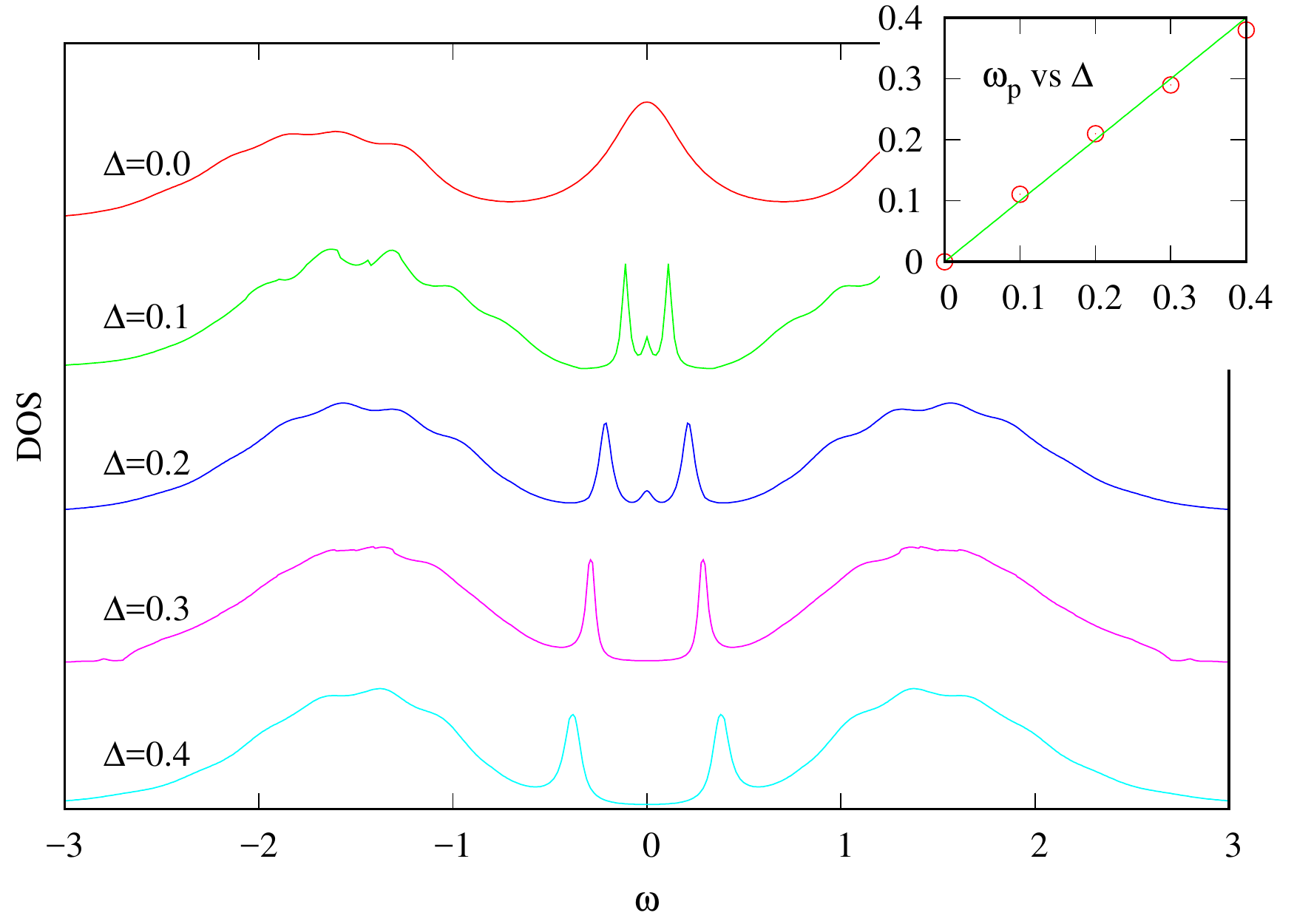}
 \caption{\label{figure3} Dependence of the position $\omega_p$ of the QP$_{W}$ peak with $\Delta$ in the narrow band for $U=3$ and $t_2/t_1=0.5$. For small $\Delta$ the system is metallic. Inset: Linear dependence of the peak positions vs $\Delta$. 
 } 
 \end{figure}

 To characterize the nature of these peaks we calculated the one-particle energy spectrum for a small system (see Supplementary material) and noticed that the largest weight in this excitation, for example, for the orbital selective Mott (OSM) phase, corresponds to states with a hole in the wide band and a doubly occupied state in the narrow band (for the positive energy QP) and viceversa (for the negative energy one), suggesting a holon-doublon bound state. To verify this we calculated the excitation spectrum $Im G^>_{W}=Im \langle a_{W} (\omega + i\eta -H_{imp})^{-1} a^{\dagger}_{W} \rangle$ where 
the operator  $a^{\dagger}_{W}=(1-n_{1\uparrow})(1-n_{1\downarrow})n_{2\downarrow}c^{\dagger}_{2\uparrow}$ acts on the ground state and creates a doubly occupied state in the narrow band when there is a hole in the wide band, thus creating a QP$_W$ (for QP$_N$ interchange 1 and 2). The Hamiltonian is provided by the converged DMFT loop for an 8 site system (S2).  In Fig.\ref{figure2} we plot the excitation spectrum of these operators (blue and green curves, arbitrary units). In this situation of OSM (left panel) we only have QP$_W$ and only $G^>_{W}$ peaks at the QP energy in the NB, verifying the holon-doublon nature of these excitations. A similar result is expected for $G^<_{W}$.  

QP peaks in one band, thus, arise when there are holes or doubly occupied states in the ground state of the other band ({\it i.e}, it is metallic). When one band becomes insulating (for example, the narrow band at the OSM point), there are no holes or two-electron states available and QP peaks will not form in the wide band. Coherent peaks in one band are correlated with QP peaks in the other. 

Previous works reported holon-doublon pairs in related models at higher energies\cite{fulde,wang,picos,karski} (also observed here close to the Hubbard bands, see Fig.\ref{figure1}a) and b)) and also as metastable states out of equilibrium \cite{prelovsek,rincon}  

However,the QPs found in our work are stable charge neutral, spin singlet, orbital triplet holon-doublon inter-orbital bound pairs which stem out as well defined peaks and can be completely separated from the Hubbard bands depending on $\Delta$. Such Hubbard excitons emerge due to the repulsive interband interaction ($U_{12}\sim U$) and are present even if $J=0$ (see detailed description in the Supplementary Material). When both bands are metallic with different bandwidths there are light and heavy QPs (QP$_W$ and QP$_W$) which become equal for equal bandwidths. The QP$_N$ vanish when the NB turns insulating. 

\begin{figure}
 \includegraphics[scale=0.45]{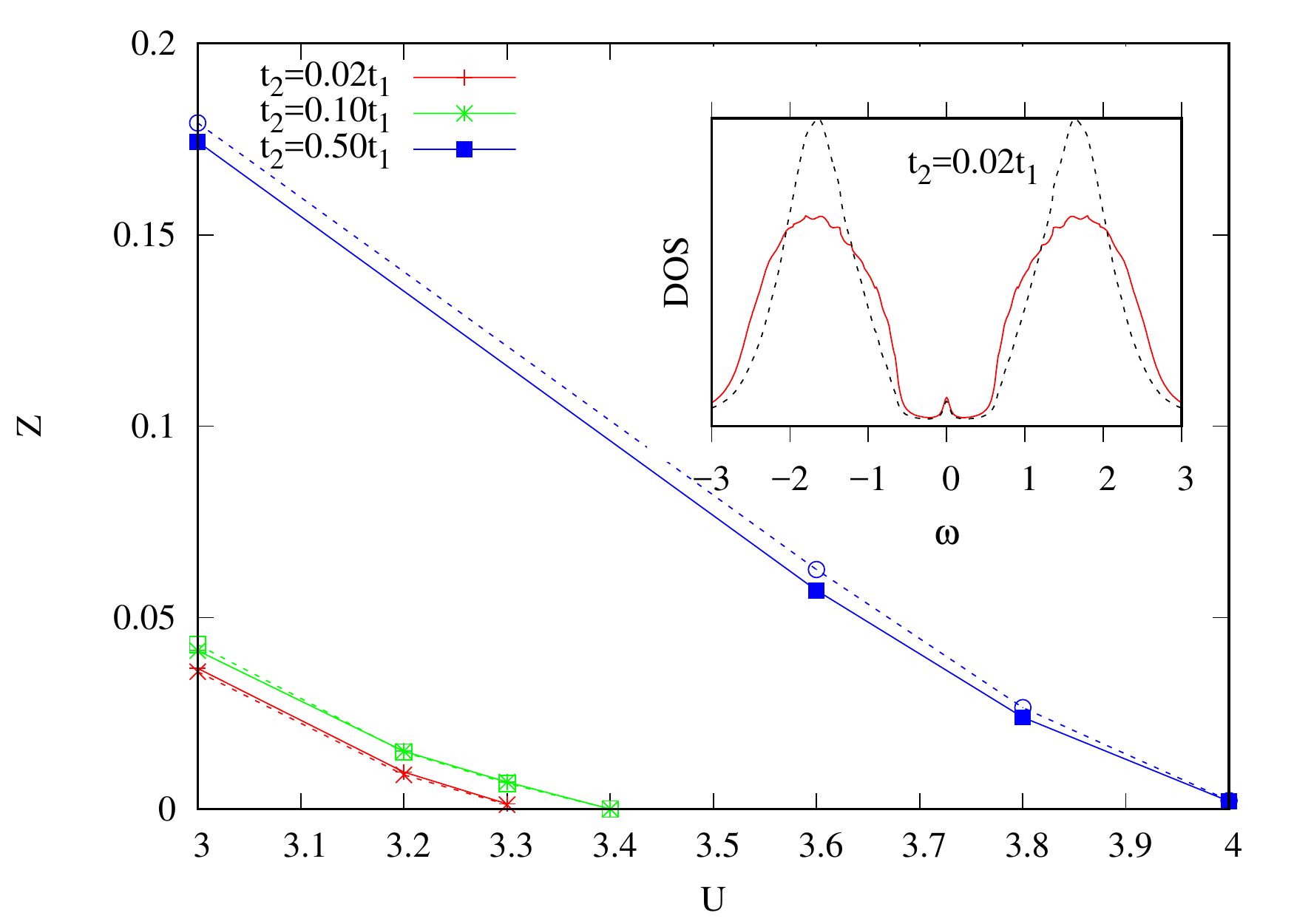}
 \caption{\label{figure4} Quasiparticle weights $Z$ in each band vs $U$ for $\Delta=0$, $t_2=0.02 t_1$, $t_2=0.1 t_1$ and $t_2=0.5 t_1$ showing a locked metal to insulator transition even for very small bandwidth ratios. Inset: DOS showing the peaks at the Fermi energy at $t_2=0.02 t_1$, close to the transition for $U=3.3$ (full red line: wide band; dotted black line: narrow band. 
 } 
 \end{figure}

{\it Rotationally invariant case $U=U_{12}$}: 
As seen in Fig. \ref{figure3}, when $\Delta=0$ (corresponding to $J=0$ in the full KH model), both QP peaks merge at the Fermi energy and the narrow band will have a finite DOS at this energy as long as the wide band is metallic. 
In fact, in this limit we do not find an OSM phase for any ratio of finite band widths. We find that for this highly symmetric interaction, both bands, even with very different widths, are "locked" to each other, using the words of Ref.\cite{KotliarYashar,si}, (see Fig.\ref{figure4}), and they transition to an insulating state simultaneously and continuously when increasing $U$. The enhanced symmetry of this case prevents the OSM quantum phase transition to take place.

In Fig. \ref{figure4} we show the dependence of the quasiparticle weights of each band $\alpha$ defined as $Z_{\alpha}^{-1}=1-d Re \left[ \Sigma_{\alpha}(\omega)\right]/d{\omega}|_{\omega=0}$ for three band ratios depicting the simultaneous transition for both bands.  From here it is clear that, for very small $t_2/t_1$ (no Kondo mechanism in band 2) the holon-doublon QP$_W$ peaks in the NB at $\omega=0$ add up to the same weight as the originating Kondo peak in the WB ($Z_1=Z_2)$. When $t_2\sim t_1$ both, Kondo and Hubbard QP mechanisms are activated in both bands, and the corresponding peaks at the Fermi energy add up to form the central peak with a larger weight (in this figure there is a small difference between the $Z_{\alpha}$ probably due to finite size effects, however, the transition is still simultaneous). As a consequence, when the bandwidth ratio grows towards 1, the critical interaction for the metal-insulator transition increases and should read the SU(4) critical value.\cite{su4} 

Also shown is the DOS for a small band ratio close to the transition where a clear finite DOS at the Fermi energy is seen for both bands. Here we also find an interesting feature in the Hubbard bands: While the wide band presents the well-known dome form for this Bethe lattice, the narrow band Hubbard bands are nearly triangular, having the same support as the wide band (caused by the interband Coulomb interaction $U_{12}$) and narrowing towards the middle trying to mimic a narrow band form. 

A very interesting feature in this extremely small band ratio case which deserves deeper further analysis is that both zero-energy peaks correspond to different kinds of excitations: pure Kondo-like spin exchange in the wide band and pure interband Hubbard holon-doublon excitons in the narrow one (QP$_W$). This difference is shown in Fig.\ref{figure2} where we calculate the ground state spin-spin correlation between the effective impurity sites corresponding to each band and their corresponding bath $\langle {\bf S}_{\alpha} \cdot {\bf s}_{bath\alpha} \rangle$. This correlation is nearly -3/4 for the Kondo singlet state in the wide band and nearly cero for the extremely narrow band, whereas the holon-doublon response function $G^>_{W}(\omega)$ peaks at zero energy in the narrow band and $G^>_{N}(\omega)$ is zero for $\omega=0$ in the wide band (right panel). There are no low-energy free holons (doublons) in the narrow band, being always bound to doublons (holons) in the wide band while free holons and doublons do exist in the metalic, wide band.

Our results are in agreement with those by \cite{Liebsch,Koga,Cornaglia} but not with \cite{Ferrero2005,demedicigeorgesbiermann,demedici}, where it was claimed that for bandwidth ratios with $t_2/t_1<0.2$ there would be an OSM transition for $U=U_{12}$, albeit, with a finite DOS close to the Fermi energy in the insulating band.\cite{Ferrero2005,demedicigeorgesbiermann} Our findings lead us to conclude that this corresponds, in fact, to structure of the central peak in the narrow band at the Fermi energy, which is split due to the approximate methods used. The results presented here shed light on this long lasting controversy. When $U_{12} \lesssim U$  (assuming that $t_2 \ll t_1$), both QP peaks overlap (as long as their width $\sim Z_1 t_1 \sim \Delta$) and there could be an OSM transition at a critical value of $U$.

{\it Conclusions:} 
By using state-of-the-art numerical calculations for the non-hybridized two-orbital Hubbard model and a repulsive inter-orbital Coulomb interaction we find a novel mechanism that causes attraction from repulsion leading to the emergence of robust quasiparticle states. They consist of inter-band holon-doublon bound states (Hubbard excitons) at energies $U-U_{12}$ which are correlated to the central coherent peak in the other band. 

As a consequence, at the symmetric point, when $U=U_{12}$, there will be a simultaneous and continuous Mott transition (no OSM phase) at a critical interaction, for any finite ratio of band widths. This means that a mechanism that causes a gap opening in the wide band will automatically produce a gap in the narrow one as well. It also implies that there could be a Mott insulator-to-metal transition in the heavy band triggered by the proximity of a metallic band (increasing $U_{12}$ towards $U$) with no hybridization between them. 

The results presented here could explain interesting in-gap features observed in materials where this model is applicable, for example, in optical conductivity measurements or photoemission spectra. This mechanisms should be taken into account for related models, like for the OSMT in momentum space and pseudo gap formation, valid for cuprate superconductors \cite{KotliarCornaglia}. These quasiparticles should be observed also in experiments in cold atoms systems.\cite{bloch}

Our results should be robust with doping, with inter-orbital hybridization and crystal field splitting, in the presence of a finite Hund coupling $J$ between the bands (where we expect the QP peaks to have more structure due to spin splitting), and also for lattices other than the Bethe lattice considered here. We leave for future studies the analysis of the dispersion of these QPs, their temperature behavior,  as well as their possible existence in related models.

\begin{acknowledgments}
We acknowledge support from projects PICT 2012-1069 and PICT 2016-0402 from the Argentine ANPCyT and PIP 2015-2017 (CONICET).  GK was supported by DOE BES under grant No. DE-FG02- 99ER45761. This work used XSEDE, which is supported by NSF grant ACI-1548562. We thank B. Alascio, C. Balseiro, P. Cornaglia, J. Facio, M. Imada and J. von Delft for useful discussions. 
\end{acknowledgments}

\newpage
\setcounter{equation}{0}
\setcounter{figure}{0}
\setcounter{table}{0}
\setcounter{page}{1}
\makeatletter
\renewcommand{\theequation}{S\arabic{equation}}
\renewcommand{\thefigure}{S\arabic{figure}}

\section{Supplementary Material}

\subsection{DMFT+DMRG Method}

We solve Hamiltonian (1) using the Dynamical Mean Field Theory (DMFT)\cite{review} in the limit of infinite dimensions on the Bethe lattice, where the lattice is mapped onto a single impurity under a self-consistent condition. We use the DMRG to obtain the mapped impurity Green's functions at each iteration of the DMFT \cite{garcia, EPLreview}, which leads to reliable results for the real energy axis directly and also for all energy scales, suffers from no sign problem and can be used at zero temperature, having important advantages when compared to other methods like NRG and CTQMC.

The DMFT equations involve the calculation of the local Green's functions given by:

\begin{equation}
G_{\alpha}(\omega)=2\left[z+\sqrt{z^{2}-4t_{\alpha}^{2}}\right]^{-1}\label{eq:latticeGreen}
\end{equation}
where $z=\omega+\mu-\Sigma_{\alpha}(\omega)$ and $\Sigma_{\alpha}(\omega)$
is the self-energy for orbital $\alpha$, $\alpha=1,2$. The local
self-energy $\Sigma_{\alpha}(\omega)$ is calculated using an auxiliary
Hamiltonian for an impurity coupled to a non-interacting bath:

\begin{eqnarray}
H_{imp} & =-\mu & \sum_{\alpha\sigma}n_{0\alpha\sigma}+\\
& + & U\sum_{\alpha}n_{0\alpha\uparrow}n_{0\alpha\downarrow}+\sum_{0\sigma\sigma'}U_{12}n_{0,1\sigma}n_{0,2\sigma'}+H_{b}.\nonumber
\label{eq:Himp}
\end{eqnarray}

The bath $H_{b}$ is:
\begin{equation}
H_{b}=\sum_{k\alpha\sigma}\lambda_{k}^{\alpha}b_{k\alpha\sigma}^{\dagger}b_{k\alpha\sigma}+\sum_{\alpha k\sigma}v_{k}^{\alpha}\left[b_{k\alpha\sigma}^{\dagger}c_{0\alpha\sigma}+H.c.\right]\label{eq:Hbath}
\end{equation}
where $b_{k\alpha\sigma}^{\dagger}$ creates an electron with spin
$\sigma$ at the bath-site $k$, associated to the orbital $\alpha$
of the impurity located at site $0$.

The algorithm is summarized as:
\begin{description}
\item [{(i)}] Start with $\Sigma_{\alpha}(\omega)=0$
\item [{(ii)}] Calculate the Green's function $G_{\alpha}(\omega)$ at
the lattice site for each orbital $\alpha$ using (\ref{eq:latticeGreen})
\item [{(iii)}] Calculate the hybridizations:
\begin{equation}
\Gamma_{\alpha}(\omega)=\omega+\mu-\Sigma_{\alpha}(\omega)-G_{\alpha}^{-1}(\omega)\mbox{.}
\end{equation}

\item [{(iv)}] Find a Hamiltonian representation $H_{imp}$ with hybridizations
$\tilde{\Gamma}_{\alpha}(\omega)$ to approximate the $\Gamma_{\alpha}(\omega)$.
$\tilde{\Gamma}_{\alpha}(\omega)$ is characterized
by the parameters $v_{k}^{\alpha}$ and $\lambda_{k}^{\alpha}$ of
$H_{imp}$ through:
\begin{equation}
\tilde{\Gamma}_{\alpha}(\omega)=\sum_{k}\frac{\left|v_{k}^{\alpha}\right|^{2}}{\omega-\lambda_{k}^{\alpha}}\mbox{.}\label{eq:star}
\end{equation}

\item [{(v)}] Calculate the Green's functions $G_{\alpha}(\omega)$ at
site ``$0$'' of the Hamiltonian $H_{imp}$ using DMRG.\cite{white1,karen1}
\item [{(vi)}] Obtain the self-energy
\begin{equation}
\Sigma_{\alpha}(\omega)=\omega+\mu-G_{\alpha}^{-1}(\omega)-\tilde{\Gamma}_{\alpha}(\omega)\mbox{.}
\end{equation}
Return to \textbf{(ii)} until convergence.
\end{description}

\subsection{The Hubbard bound states}

\subsection{The quasiparticles}

To understand the nature of these quasiparticles we have diagonalized
a small Hamiltonian including both impurities and a bath site for
each (zero bandwidth limit) which mimic the full DMFT solution. After
showing that this small system captures the physics of the novel quasiparticle
states, we proceed to its detailed analysis.

\begin{figure}[h]

 \includegraphics[scale=0.45]{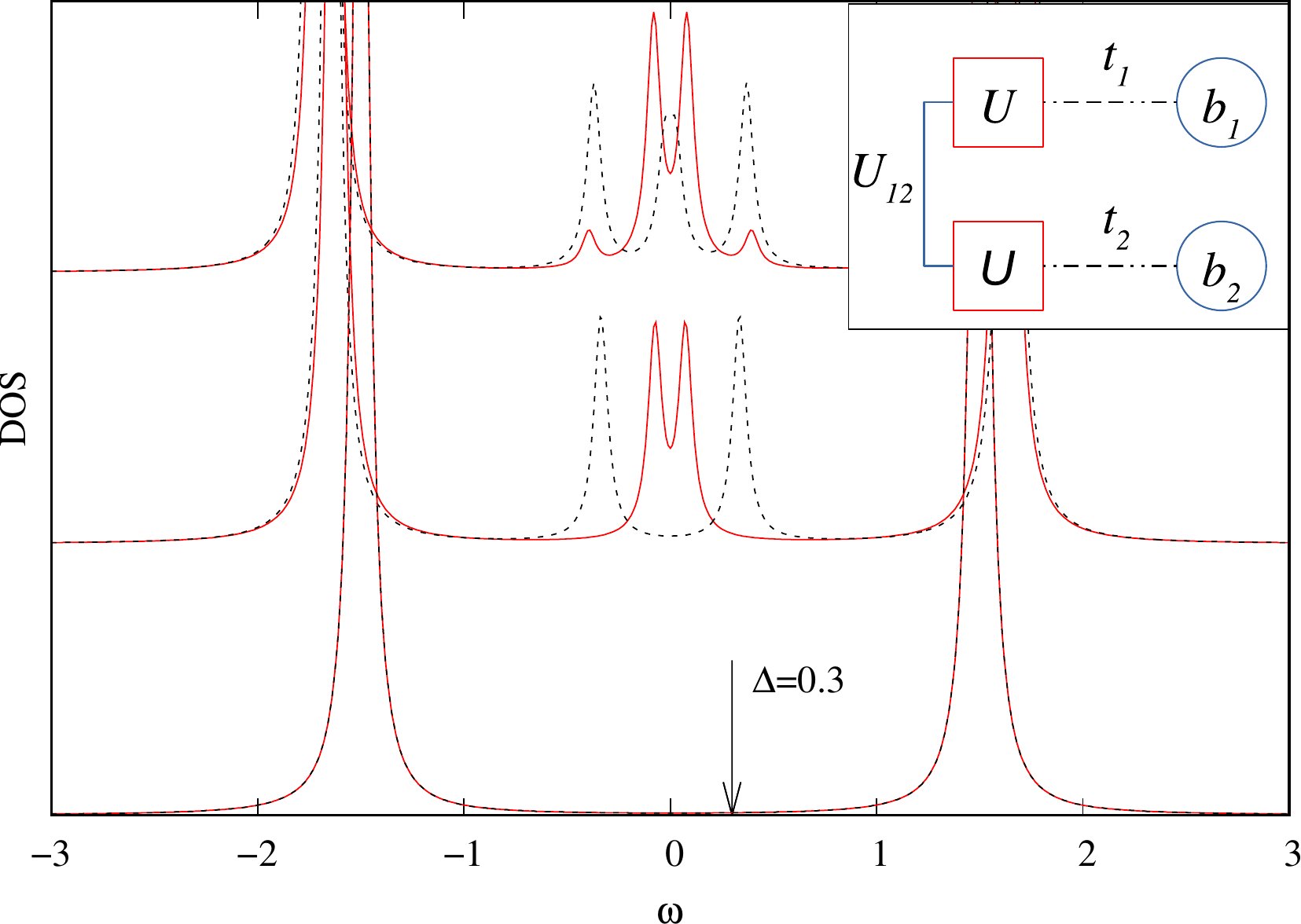}
 \caption{\label{figures2}  The DOS for the fully metallic ($t_1=0.2, t_2=0.1)$, OSM ($t_1=0.2, t_2=0.001$) and fully insulting states ($t_1=0.002, t_2=0.001$), from top to bottom for $U=3$. Red (black) lines correspond to the wide (narrow) "band". The central Kondo-like peak is split due to the finiteness of the system. The QP peaks are located at energy $\Delta=0.3$ and correspond to holon-doublon bound pairs. Inset: Sketch of the small system calculation with exact diagonalization to interpret the DMFT calculations: Squares represent the impurities and circles the bath sites.
 } 
 \end{figure}

The eigenstates of the Hamiltonian can be written in terms of a basis
$|i_{1}\rangle|i_{2}\rangle|b_{1}\rangle|b_{2}\rangle$, where the
first and second indices ( $i_{1}$ and $i_{2}$) represent the impurity
orbitals of the DMFT equations and the third and fourth indices are
the corresponding states of the bath (see the inset of Fig. \ref{figures2}).
The indices run over up, down, empty and doubly occupied states $\{0,\uparrow,\downarrow,\uparrow\downarrow\}$.

We fine tuned the ``Mott'' transition by changing the hopping parameters
$t_{1}$, $t_{2}$, while keeping $U=3$ and $\Delta=0.3$. Fig. \ref{figures2}
replicates in a very simple way our DMFT results for the OSMT, with
delta functions at the peak positions (note the split structure for
the central coherent peak stemming from the finiteness of the system,
\cite{Hewson}).

If we define the ground state as:

\[
\left|\Phi_{0}\right\rangle =\sum_{i_{1},i_{2},b_{1},b_{2}}A_{i_{1}i_{2}b_{1}b_{2}}|i_{1},i_{2}\rangle|b_{1},b_{2}\rangle ,
\]
we can specify the relevant states for each case (metallic, OSM and insulating) for $N=4$ particles (half-filled) and energy $E_{0}$ and the parameters in Fig \ref{figures2}:

\begin{widetext}

\[
\left|\Phi_{0}\right\rangle \approx\begin{cases}
\alpha\sum_{\sigma\sigma'}|\sigma,\sigma'\rangle|\bar{\sigma},\bar{\sigma'}\rangle+\beta_{1}\sum_{\sigma.x}|x,\sigma\rangle|\bar{x},\bar{\sigma}\rangle+\beta_{2}\sum_{\sigma.x}|\sigma,x\rangle|\bar{\sigma},\bar{x}\rangle+\gamma\sum_{x}|x,\bar{x}\rangle|\bar{x},x\rangle & \mbox{metallic }\\
\alpha\sum_{\sigma\sigma'}|\sigma,\sigma'\rangle|\bar{\sigma},\bar{\sigma'}\rangle+\beta_{1}\sum_{\sigma.x}|x,\sigma\rangle|\bar{x},\bar{\sigma}\rangle & \mbox{OSM }\\
\alpha\sum_{\sigma\sigma'}|\sigma,\sigma'\rangle|\bar{\sigma},\bar{\sigma'}\rangle & \mbox{insulating }
\end{cases}
\]
\end{widetext}
where $\sigma=\uparrow, \downarrow$ and $x=0, \uparrow\downarrow $. The coefficients are $\alpha=\{-0.471;-0.485;0.5\}$, $\beta_{1}=\{0.123;0.121;0\}$, $\beta_{2}=\{-0.124;0;0\}$, $\gamma=\{0.074;0;0\}$, corresponding
to \{metallic, OSM, insulating\}, respectively.

We now introduce the state $\left|K_{1}\right\rangle $ with $N=5$
particles and energy $E_{0}+\omega_{1}^{K}$, such that $\left\langle K_{1}\right|c_{1\downarrow}^{\dagger}\left|\Phi_{0}\right\rangle $
is responsible for the small system analog of the Kondo resonance
at $\omega=\omega_{1}^{K}\sim0$ in band 1. The principal term
for the state is $\left|K_{1}\right\rangle \approx\alpha_{1}^{K}\sum_{\sigma}|\downarrow,\sigma\rangle|\uparrow\downarrow,\bar{\sigma}\rangle$,
which stems from the term $\beta_{1}\sum_{\sigma.x}|x,\sigma\rangle|\bar{x},\bar{\sigma}\rangle$
in the ground state. The numerical coefficients are $\alpha_{1}^{K}=\{0.689;0.701;0.707\}$,
and $\omega_{1}^{K}=\{0.08;0.074;0\}$. Similarly, $\alpha_{2}^{K}=\{0.68;0.686;0.707\}$,
and $\omega_{2}^{K}=\{0.027;0;0\}$ for the Kondo resonance at orbital
2.

The new bound wide quasiparticle QP$_W$ (QP peak in the NB) can be explained by introducing the
state $\left|W\right\rangle $ with $5$ particles and energy $E_{0}+\omega^{W}$,
such that $\left\langle W\right |c_{2\downarrow}^{\dagger}\left|\Phi_{0}\right\rangle $
is responsible for the peak at $\omega=\omega^{W}\sim\Delta$ in
band 2. Roughly, $\left|W\right\rangle \approx\alpha^{W}|0,\uparrow\downarrow\rangle|\uparrow\downarrow,\downarrow\rangle$
which stems from the term $\beta_{1}\sum_{\sigma.x}|x,\sigma\rangle|\bar{x},\bar{\sigma}\rangle$
in the ground state. Notice that $\left|W\right\rangle $ shares its
origin with $\left|K_{1}\right\rangle $, explaining their similar weigth
in our DMFT calculations. The numerical coefficients are $\alpha^{W}=\{0.969;0.975;1\}$,
and $\omega^{W}=\{0.37;0.336;0.3\}$. 

The equivalent narrow quasiparticle, QP$_N$,  (QP peak in the WB) $\left|N\right\rangle $, with $5$ particles and energy $E_{0}+\omega^{N}$,
originates from $\left\langle N\right |c_{1\downarrow}^{\dagger}\left|\Phi_{0}\right\rangle $
which gives rise to the peak at $\omega=\omega^{N}\sim\Delta$ in
band 1. Roughly, $\left|N\right\rangle \approx\alpha^{N}|\uparrow\downarrow, 0\rangle|\downarrow,\uparrow\downarrow\rangle$
which stems from the term $\beta_{2}\sum_{\sigma.x}|\sigma,x\rangle|\bar{\sigma},\bar{x}\rangle$
in the ground state.
 The numerical coefficients are $\alpha^{N}=\{0.976;0.987;1\}$, and $\omega^{N}=\{0.394;0.368;0.3\}$.

We omit here the detailed characterization of the Hubbard bands, see
for instance \cite{Hewson}.

\begin{figure}

 \includegraphics[scale=0.8]{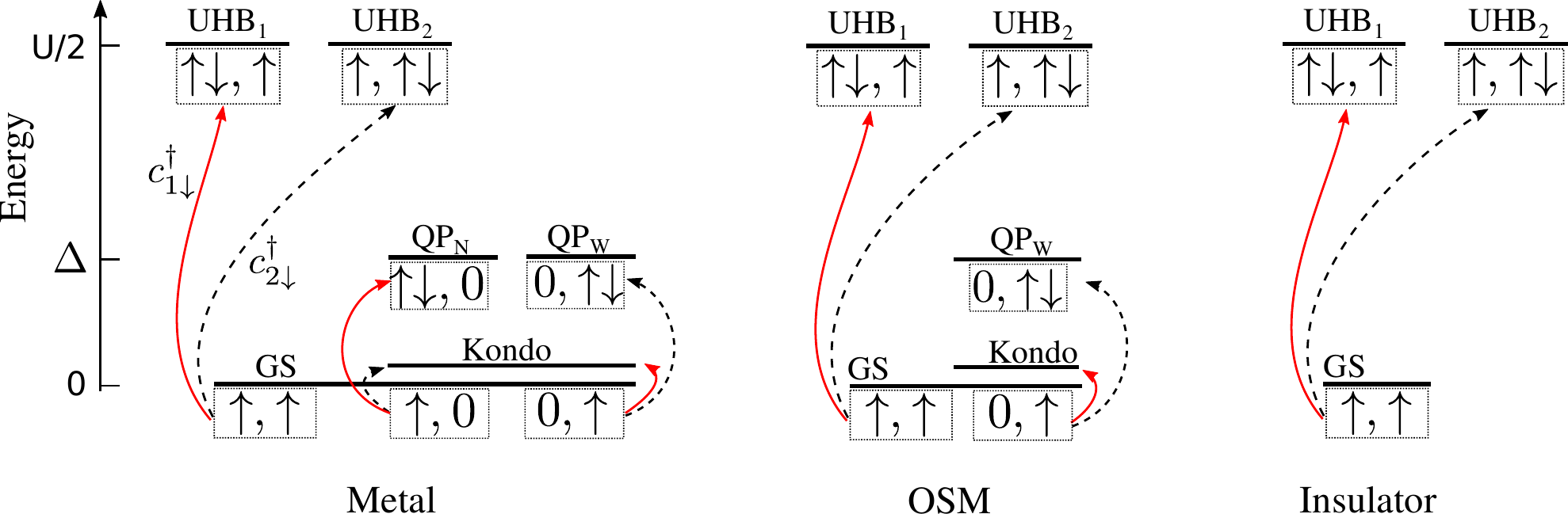}
 \caption{\label{sketch}  Sketch of the relevant states of the effective impurity $|$WB,NB$\rangle $ (omitting the bath sites, for clarity) participating in each main feature of the DOS: the central Kondo peak, the in-gap QP peaks and the Hubbard Bands (here only positive energies are shown): a) Fully metallic, b) OSM: Metal in WB/Insulator in NB, c) Fully insulating. Note the color codes for the creation operators. Single up spins can be also down spins. 
 } 
 \end{figure}

In conclusion, the well known low energy coherent peak in the metallic
states has charge and spin fluctuations, having an appreciable density
of holes and doubly occupied sites with a large weight in states of
the form $|\sigma,\sigma'\rangle$ (showing only the impurity configurations),  involved in the Kondo-like fluctuations.
The states in the upper Hubbard ``bands'' (extremely narrow here)
can be explained as an itinerant doubly occupied states on a sea of
singly occupied sites of the form $|d,\sigma\rangle$ and $|\sigma,d\rangle$
(and similarly for the lower Hubbard band, with holes instead of ``$d$''
configurations). Finally, the states corresponding to the low-lying QP peaks have a large weight in states of the form $|0,d\rangle$ or $|d,0\rangle$, that is, a neutral holon-doublon pair between the bands with energy $U$ in the atomic limit and differing in energy $\Delta=U-U_{12}$ when compared to states of the form $|\sigma,\sigma'\rangle$ with energy $U_{12}$ that have the largest weight in the ground state. In Fig. \ref{sketch} we sketch the relevant states and procedures that give rise to the different main excitations (Kondo, QPs and Hubbard bands) for the fully metallic, OSM and fully insulating states.

\widetext

\end{document}